\documentclass[epsfig,12pt]{article}
\usepackage{epsfig}
\usepackage{graphicx}
\usepackage{array}
\usepackage{color}
\usepackage{bm}
\usepackage{amsmath}%

\newcommand{\beq}{\begin{equation}}   
\newcommand{\eeq}{\end{equation}}
\newcommand{\beqn}{\begin{eqnarray}}   
\newcommand{\eeqn}{\end{eqnarray}}

\newcommand{\gsim}{\lower.7ex\hbox{$
\;\stackrel{\textstyle>}{\sim}\;$}}
\newcommand{\lsim}{\lower.7ex\hbox{$
\;\stackrel{\textstyle<}{\sim}\;$}}
\begin{document}

\begin{flushright}
FTPI-MINN-19/07,  UMN-TH-3816/19
\end{flushright}

\begin{center}
{  \large \bf  Teaching Physics at School and Colleges\,\footnote{Invited contribution to the inaugural issue of {\em Physics Educator.}}}

\vspace{2mm}

Mikhail Shifman

{\small Ida Cohen Fine Professor of Theoretical Physics}

\vspace{1mm}

William I. Fine Theoretical Physics Institute, 
University of Minnesota

116 Church Street SE, Minneapolis, MN 55455, USA

\end{center}

\vspace{3mm}

\begin{flushright}\framebox{
 \begin{minipage}[t]{22em}
{\small\em 
[Sometimes] I read in a newspaper, ``The scientist says that this discovery may have importance in the cure of cancer." The paper is only interested in the use of the idea, not the idea itself. Hardly anyone can understand the importance of an idea, it is so remarkable. Except that, possibly, some children catch on. And when a child catches on to an idea like that, we have a scientist. These ideas do filter down, and lots of kids get the spirit -- and when they have a spirit you have a scientist. It's too late for them to get the spirit when they are in our universities, so we must attempt to explain these ideas to children. }
-- Richard Feynman \cite{feyn}

\end{minipage}}
\end{flushright}

\vspace{5mm}

When I meet new people and tell them that I am a physicist they respond -- almost always with awe --  ``Oh, that's something terribly complicated'' or ``I know nothing about physics, it was the hardest discipline for me at school.''

Physics is a part of the high school science curriculum, along with biology and chemistry. As a rule, school students are allowed to choose only two out of the three disciplines, and many prefer to skip physics altogether because of its reputation as a difficult science. For the majority of school graduates this choice is not unreasonable -- they will hardly ever need physics in their adult life. Still, a significant number of graduates, those who aim at colleges, will have to deal with physics at various levels. Courses of calculus-based physics are mandatory for all future engineers, pre-med students and in some instances students of chemistry, let alone those who chose physics as their major.

As a professor at the University of Minnesota I taught such courses for a number of years -- some of the classes' enrollment was quite high, 200 students or more -- and was always surprised by how few of them managed to earn good grades, despite their previous exposure to physics at high school. For pre-med students low grades in physics meant the end of their medical career before it even started. 

My first guess was that, perhaps, the corresponding textbooks were far from perfection. Having inspected a few of the most popular school textbooks I found them satisfactory, both with regards to the coverage of relevant topics and exercise problems. 

All these textbooks have one common feature: they present 1000 (or more) page volumes -- a huge reservoir of knowledge which is supposed to be consumed in one school year. It is easy to understand that, even if physics classes take place every day of the week, a student will have to digest 6-7 pages a day. If you add a few problems which are absolutely necessary to develop at least a minimal understanding of material the task becomes impossible. The rate of learning becomes too high and not sustainable. My own observations show that an average student after going through this standard routine at school either has huge gaps, or acquires only superficial knowledge which usually leads to failure on quizzes and examinations. This, I believe, is a true reason why entry level university courses in physics present such a difficult barrier for the majority of students aspiring to overcome it.

Below I will suggest a remedy, at least partial, but first I want to tell how, to my mind, physics could be taught at school in an ideal world. As an example, I will take my own school in Moscow of the 1960s. If there were any positive aspects of life in the USSR at all, the role of physics in school curricula was one of them. The course of physics spanning mechanics, thermodynamics, electrostatics, electricity and magnetism, optics and the basics of special relativity, lasted for five years, with two lessons a week. 
Note that at that time the school week in the USSR included Saturdays. Thus, students had three days to think over and digest the contents of each lesson and solve one or two problems. 

Problem-solving is an extremely important element of education in physics. It is as important (if not more) than lectures in classes. Without systematic exercising no  understanding of physics can be achieved. The narratives that students hear in classes do not anchor in their minds and usually fade away rather quickly being superseded by subsequent topics. In this sense physics education is no different from, say, music education. 

Every few weeks solutions of sample problems were discussed in class, with the best solutions being presented by their authors for public debate. The above routine proved to be optimal, resulting in a long-lasting effect. Needless to say, it is hardly possible to teach in this mode under the five-days-a-week regime.

I realize that the ``ideal'' pattern I have described above cannot be implemented in the US. Families often move from town to town, children change schools, and it is impossible to synchronize curricula all over the country if a course lasts for five years. Dense curricula with a high rate of presentation of new material (the five-days-a-week regime) is a practical necessity. What can be done to alleviate this situation in physics education and help future college students to succeed in physics courses? How does one captivate their imagination turning problem solving into something entertaining?

To my mind, a magazine published in Russia and entitled {\em Kvantik} \cite{kv} (Little Quantum) can serve as a guideline. It is released monthly and intended for children starting from the age of 12-13. The most important difference from many popular books on science for children existing in the market is its continuity: every issue of {\em Kvantik} carries funny age-appropriate stories related to math and physics, colorful pictures and lots of problems at various levels, with solutions discussed in the subsequent issues. Some stories are of to-be-continued type and could be presented from different angles. Reading this magazine is not a one-time affair, rather it is an extended process which can last four or five years, providing future college students with sufficient background and, most importantly, teaching them to strive in problem solving on a regular basis. One can find there articles and problems to any taste.

\begin{figure}[h]
\begin{center}
\includegraphics[width=4in]{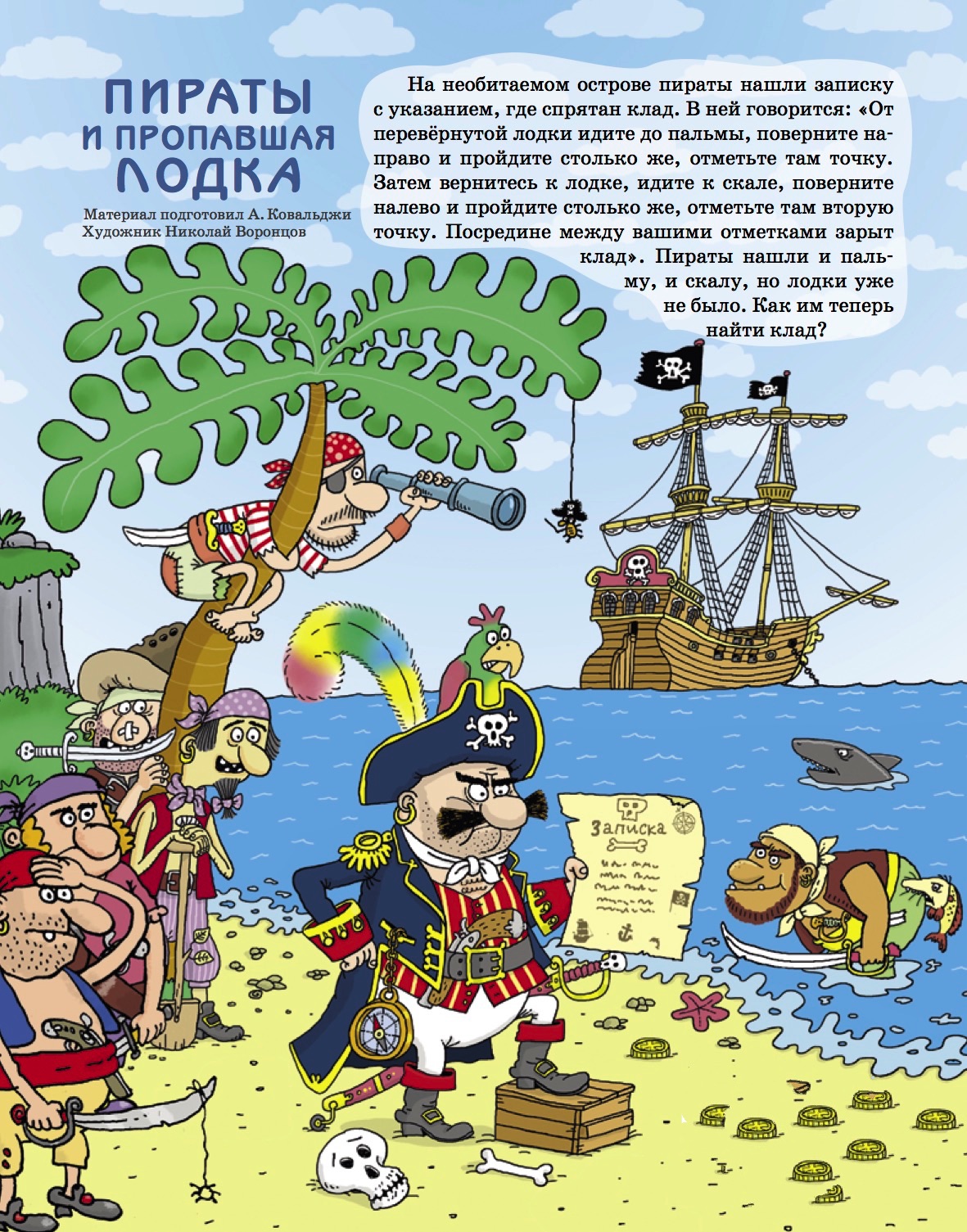}
 \end{center}
 \caption{\small This problem due to A. Kovaldzhi was published in the December 2018 issue of {\em Kvantik}. Illustration by N. Vorontsov.}
 \label{pirates}
 \end{figure}

As an example I will present below a curious problem which can be solved by anyone with basic knowledge of planar geometry. The problem illustrated in Fig. \ref{pirates} is as follows. Pirates found a piece of parchment on an uninhabited island about the location of a treasure trove which had been hidden somewhere on the island by their predecessors. The note on the parchment read:

\begin{quote}
Start from the capsized boat. Go in the direction of the palm tree carefully counting the number of steps. When you reach the palm tree turn exactly right and make exactly the same number of steps. Mark the point you arrived at. Then return to the boat. Go in the direction of the rock counting the number of steps. When you reach the rock
turn left. After having made the same number of steps mark the second point. The treasure trove  is in the middle of the line connecting two marked points.

\end{quote}

 The problem is that the capsized boat was nowhere in sight, it had disappeared.  How can the pirates  still find the treasure trove?

 \newpage

\begin{figure}[h]
\begin{center}
\includegraphics[width=5.3in]{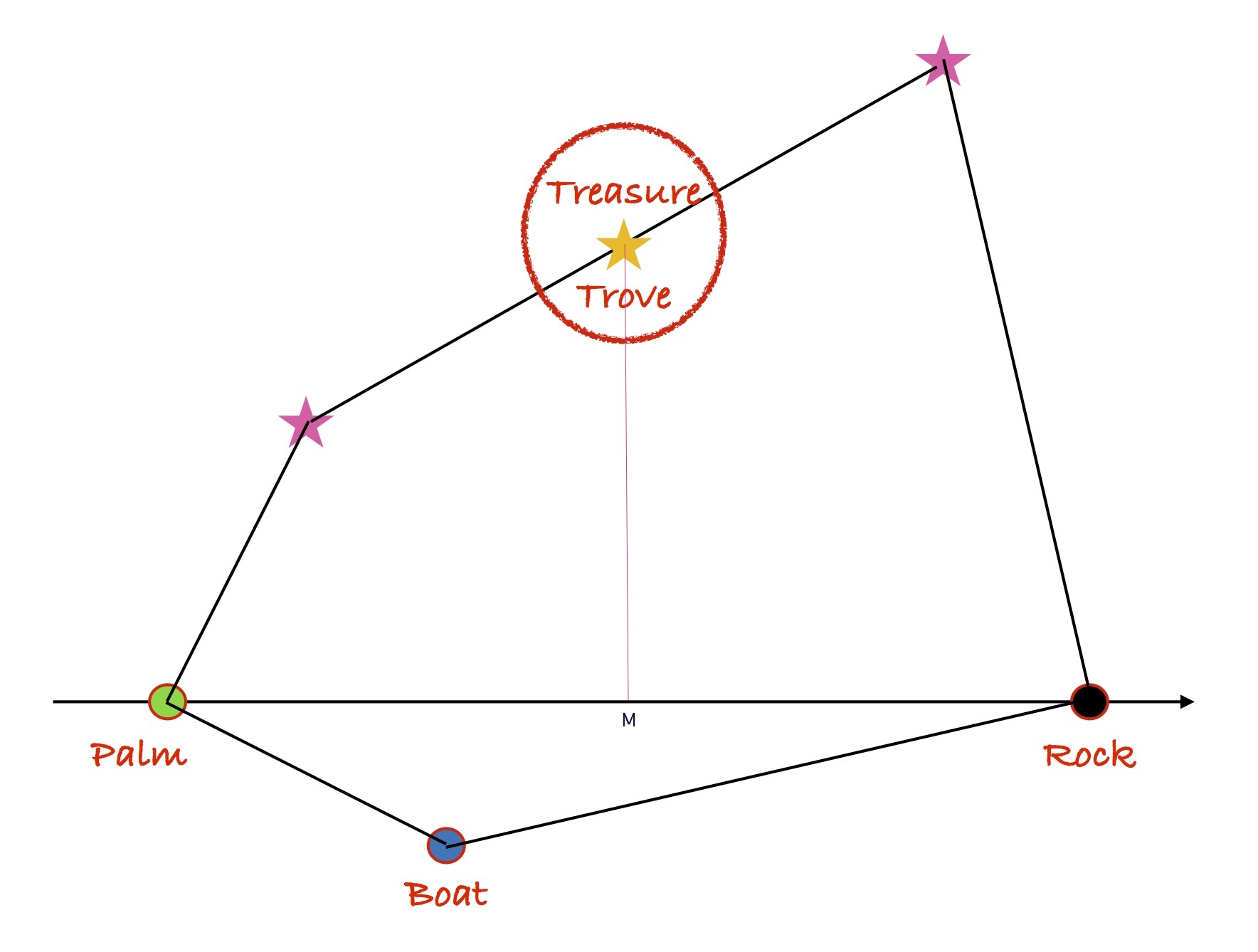}
 \end{center}
 \vspace{-6mm}
\caption{\small The map I draw to indicate the position of the treasure trove.}
 \label{plan}
 \end{figure}

 To solve the problem let us first use our physical intuition. The problem is formulated in such a way that it informs us that the solution exists. In the absence of the boat mentioned in the parchment the only physical information at our disposal is the line $PR$ connecting the Palm tree and the Rock (see Fig.~\ref{plan}). Due to the left-right symmetry
 it is quite obvious that the treasure trove (marked by a yellow star) is located on the line perpendicular to $PR$ crossing the latter exactly in the middle (see the point marked M). This result does not depend at all on the location of the disappeared boat. 
 
This is as far as we can go with qualitative statements. To prove that the length of the perpendicular line is exactly half that of the horizonatal line $PR$
one must draw a number of lines forming certain  triangles in Fig. \ref{plan} and compare the triangles. I leave this exercise to the reader.

\newpage

\section*{Acknowledgments}
I an grateful to Daniel Schubring for useful discussions. This work is supported in part by DOE grant DE-SC0011842.


\begin{thebibliography} {99}



{\small

\bibitem{feyn}
Richard Feynman, ``The Value of Science,'' {\em Engineering and Science}, Vol. 19, December 1955 Issue, page 13.

\bibitem{kv}
{\em Kvantik}, https://kvantik.com (in Russian). {\em Kvantik}'s older ``brother,'' the magazine {\em Kvant} intended for curious young people age 17 and older also exists.
On webcite http://kvant.mccme.ru one can find its archive including all issues from January 1970 till present.
}

\end{thebibliography}
\end{document}